\documentclass[sigconf]{acmart}

\AtBeginDocument{%
  \providecommand\BibTeX{{%
    \normalfont B\kern-0.5em{\scshape i\kern-0.25em b}\kern-0.8em\TeX}}}

\renewcommand\footnotetextcopyrightpermission[1]{} 		
\setcopyright{none}

\usepackage{url,hyperref,lineno,microtype}
\usepackage{array, caption, tabularx,  ragged2e,  booktabs, graphicx, blindtext, longtable, multirow, titlesec, subfigure}
\usepackage[english]{babel}
\usepackage{tcolorbox}

\begin{document}
\renewcommand{\abstractname}{PREAMBLE}

\title[Fairness in Tourism Recommender Systems]{Towards Individual and Multistakeholder Fairness in Tourism Recommender Systems}

\author{Ashmi Banerjee}
\email{ashmi.banerjee@tum.de}
\affiliation{%
  \institution{Technical University of Munich}
  \streetaddress{Boltzmannstrasse 3}
  \city{Garching}
  \country{Germany}
  \postcode{85748}
}

\author{Paromita Banik}
\email{paromita.banik@tum.de}
\affiliation{%
  \institution{Technical University of Munich}
  \streetaddress{Boltzmannstrasse 3}
  \city{Garching}
  \country{Germany}
  \postcode{85748}
}

\author{Wolfgang Wörndl}
\email{woerndl@in.tum.de}

\affiliation{%
  \institution{Technical University of Munich}
  \streetaddress{Boltzmannstrasse 3}
  \city{Garching}
  \country{Germany}
  \postcode{85748}
}

\renewcommand{\shortauthors}{Banerjee, et al.}

\begin{abstract}

At the sixth FAccTRec Workshop on Responsible Recommendation (FAccTRec '23)~\footnote{https://facctrec.github.io/facctrec2023/}, we presented and discussed the findings of our recent literature review titled "A Review on Individual and Multistakeholder Fairness in Tourism Recommender Systems." This work has been originally published in Frontiers in Big Data.
~\footnote{
  Ashmi Banerjee, Paromita Banik, and Wolfgang Wörndl. 2023. A Review on
  Individual and Multistakeholder Fairness in Tourism Recommender Systems.
  Frontiers in Big Data 6 (2023) https://doi.org/10.3389/fdata.2023.1168692}


\end{abstract}


\maketitle
\pagestyle{plain} 
\section{Motivation}

This position paper summarizes our published review~\cite{ashmiFrontiers2023} on individual and multistakeholder fairness in Tourism Recommender Systems (TRS).
Recently, there has been growing attention to fairness considerations in recommender systems (RS). It has been acknowledged in research that fairness in RS is often closely tied to the presence of multiple stakeholders, such as end users, item providers, and platforms, as it raises concerns for the fair treatment of all parties involved~\cite{AbdollahpouriBurker}. Hence, fairness in RS is a multi-faceted concept that requires consideration of the perspectives and needs of the different stakeholders to ensure fair outcomes for them. However, there may often be instances where achieving the goals of one stakeholder could conflict with those of another, resulting in trade-offs~\cite{Jannach}.

As pointed out by~\citet{10.1145/3511047.3538031}, fairness definitions often depend on various domains and contexts.
Hence, When studying fairness in multistakeholder recommender systems, it becomes crucial to identify the stakeholders who should be treated fairly, quantify potential harms, and analyze metrics that can measure and minimize these harms~\cite{ekstrand2022fairness}.
However, it is important to acknowledge that defining a metric for fairness has inherent limitations and may involve trade-offs. These limitations and trade-offs do not invalidate the concept of fairness itself, as all fairness constructs come with their own limitations and trade-offs, and there is no universally accepted definition of fairness~\cite{ekstrand2022fairness,narayanan2018translation}. 
In this context, \textit{a multistakeholder recommender system is considered to be fair if it minimizes any bias or circumstance that may result in disfavored outcomes for each stakeholder.} This implies that a fair multistakeholder RS may have to consider trade-offs in the respective stakeholder concerns.

Our work~\cite{ashmiFrontiers2023} uses analogous terminology as~\citet{AbdollahpouriBurker} to demonstrate the close connection between multistakeholder recommendation and multi-sided fairness. 
We classify fairness in TRS into four groups--- 
\begin{itemize}
    \item \textit{C-Fairness}: focuses on the \textit{consumers} or end-users who receive or want to receive recommendations to plan their trips, such as tourists, business
    travelers, airline passengers, etc. It includes \textit{individual} and \textit{group discrimination}.
    \item \textit{I-Fairness}: addresses the \textit{item providers} or the entities that provide the consumers with the recommended facility for their trips, such as hotels, resorts, rentals, amusement parks, airlines, tour operators, and vacation companies. It deals with \textit{popularity bias} and \textit{exposure bias}.
    \item \textit{P-Fairness}: concentrates on the \textit{platform} operator which hosts the
    recommender system and constitutes the \textit{ranking bias}.
    \item \textit{S-Fairness}: considers the impact on \textit{society} through \textit{responsible recommendations}. Here, the society represents the environment and entities or groups affected by the tourism activity but are not directly part of the TRS. This includes the local environment, city authorities, municipal councils, local businesses, and Destination Management Organizations (DMOs).
\end{itemize}


Each stakeholder has a vested interest in the traveler's journey, and optimizing recommendations to enhance the consumer experience can yield benefits for all parties involved~\cite{abdollahpouri2020multistakeholder}.
Nevertheless, there are situations where achieving one stakeholder's objectives may conflict with another's goals, resulting in trade-offs~\cite{Jannach}. To ensure fairness within TRS, adopting a multistakeholder approach that acknowledges the interdependence between stakeholders and the need to balance their objectives is vital.

\section{Literature Overview}
Although there have been some research efforts on multistakeholder fairness in TRS~\cite{weydemann2019defining, shen2021sar, wu2021tfrom, rahmani2022unfairness}, limited attention has been given to generating responsible recommendations that prioritize society as a stakeholder. 
In our study~\cite{ashmiFrontiers2023}, we analyzed individual stakeholder groups, providing an intra-stakeholder perspective on fairness and identifying their key fairness criteria. 
Additionally, we investigated the presence of overlapping stakeholder scenarios, revealing that several studies have employed a multi-objective optimization framework. 
This framework aims to simultaneously address fairness concerns for multiple stakeholders when generating tourism recommendations~\cite{weydemann2019defining, shen2021sar, wu2021tfrom, rahmani2022unfairness}.
Furthermore, our findings indicate that although there have been some research efforts on multistakeholder fairness in TRS~\cite{weydemann2019defining, shen2021sar, wu2021tfrom, rahmani2022unfairness}, limited attention has been given to generating responsible recommendations that prioritize society as a stakeholder. 

To address this gap, this paper introduces the concept of~\textit{Societal Fairness}, or S-Fairness, within the context of TRS. S-Fairness considers tourism's impact on non-participating stakeholders (society), such as residents who may experience increased housing prices, environmental pollution, and traffic congestion. 
Thus, to alleviate these and other issues, it is crucial to design a TRS that generates responsible recommendations by considering the interests of all stakeholders. Such systems should advocate for sustainable tourism practices and promote responsible tourism while providing personalized recommendations to users.

During our narrative literature review~\cite{ashmiFrontiers2023}, we comprehensively examined the current research on fairness in TRS. Sixty-six papers were identified and categorized into TRS and non-TRS groups based on their domain focus. For TRS, we systematically analyzed the selected papers, considering four key aspects: the fairness criteria or biases targeted, proposed solutions, evaluating results, and analyzing datasets.
First, we identified the different stakeholders involved in the recommendation process for a TRS and the primary fairness criteria for each of them. Then we examined relevant literature falling under those criteria. Next, we analyzed works that focus on multiple stakeholders simultaneously.
Finally, we also examined the following challenges of designing fair and balanced tourism recommender systems and explored possible solutions for mitigating these challenges. 
\begin{itemize}
    \item While research has been done on fairness in RS in other domains, the domain of travel and tourism remains largely unexplored. Most studies in the tourism sector have centered on fairness in accommodation and restaurant recommendations, while other areas, such as fair trip planning and transportation, have received limited attention.
    \item In particular, the concept of generating responsible recommendations or S-Fairness in tourism has been largely overlooked in the literature, despite its importance in balancing the challenges of over and undertourism and reducing the environmental impact of tourism activities. 
    \item While context-aware recommendation models~\cite{zheng2014cslim, zheng2016user} may improve the quality of recommendations, it is important to evaluate the effectiveness of multistakeholder recommendations in different contextual and temporal situations.
    \item Research on explaining recommendations with a multistakeholder fairness objective in the tourism industry is limited.
    \item Lack of availability of detailed and representative data
    \item While most studies evaluate their models through offline analysis or using existing datasets, there is a lack of focus on user acceptance of the re-ranked or fair recommended results.
\end{itemize}

In this position paper, we emphasized addressing the unique challenges of ensuring fairness in RS within the tourism domain. We aimed to discuss potential strategies for mitigating the aforementioned challenges and examine the applicability of solutions from other domains to tackle fairness issues in tourism. By exploring cross-domain approaches and strategies for incorporating S-Fairness, we can uncover valuable insights and determine how these solutions can be adapted and implemented effectively in the context of tourism to enhance fairness in RS. 

\bibliographystyle{ACM-Reference-Format}
\bibliography{references}

\vspace{30pt}
\begin{tcolorbox}[colback=black!5!white,colframe=black!5!white]
  For the full paper, please refer to the accepted version published in Frontiers in Big Data.
  \vspace{10pt}

  \textit{Ashmi Banerjee, Paromita Banik, and Wolfgang Wörndl. 2023. A Review on
  Individual and Multistakeholder Fairness in Tourism Recommender Systems.
  Frontiers in Big Data 6 (2023) https://doi.org/10.3389/fdata.2023.1168692}

\end{tcolorbox}

\end{document}